\documentclass[sigconf,nonacm]{acmart}
\usepackage{orcidlink}
\setcitestyle{numbers,sort&compress}
\AtBeginDocument{%
  }

\acmConference[Pre-print]{Make sure to enter the correct
  conference title from your rights confirmation email}{Under Review}{2025}

\begin{document}

\title{InsightEdu: Mobile Discord Bot Management and Analytics for Educators}

\author{Mihail Atanasov\orcidlink{0009-0007-3446-3248}}
\email{mihail.atanasov@tum.de}
\affiliation{
  \institution{Technical University of Munich}
  \country{Germany}}
\author{Santiago Berrezueta-Guzman\orcidlink{0000-0001-5559-2056}}
\email{s.berrezueta@tum.de}
\affiliation{%
  \institution{Technical University of Munich}
  \country{Germany}}
  \author{Stefan Wagner\orcidlink{0000-0002-5256-8429}}
\email{stefan.wagner@tum.de}
\affiliation{%
  \institution{Technical University of Munich}
  \country{Germany}}

\renewcommand{\shortauthors}{Atanasov et all}

\begin{abstract}
Modern educational environments increasingly rely on digital platforms to facilitate interaction between students and educators. Discord has emerged as a popular communication platform in academic settings, offering a combination of messaging and support for \textit{chatbot} development. However, most existing Discord bots lack specialized educational functionalities and mobile-friendly interfaces, limiting their effectiveness for instructional use. This paper presents InsightEdu, a mobile-first interaction design and reference architecture for instructor-operated Discord bots. 
The system enables educators to conduct surveys, collect feedback, and track attendance through an intuitive mobile interface. The architecture combines a SwiftUI-based iOS client application with a Python-based Discord bot server. User evaluations with educators demonstrated significant usability improvements over traditional Discord interfaces, with 92\% of participants (n = 20) reporting enhanced efficiency in managing educational interactions. This study demonstrates that mobile-first, instructor-friendly design can significantly enhance the utility of existing communication platforms for academic purposes.
\end{abstract}




\maketitle

\section{Introduction}

Interactive learning environments are increasingly essential in computer science education, particularly in large-scale and hybrid courses \cite{interactiveLearning}. While several Learning Management Systems (LMS) support content delivery, they often lack immediate in-situ mechanisms for gauging student engagement \cite{moodle}. To address this gap, educators have adopted digital communication platforms, most notably Discord, for immediate, context-specific interaction due to its flexible channel structure, real-time messaging, and support for custom bots \cite{berrezueta2024interactive}. Such platforms can help cultivate transferable soft skills that correlate with early academic success \cite{andree2025softskills}. However, Discord’s general-purpose design does not natively support educational management tasks such as attendance tracking, feedback collection, and survey administration \cite{wulanjani2018discord}.

Extending Discord with bots is promising. However, current solutions present notable barriers: reliance on cumbersome text-based command interfaces for instructors, limited mobile accessibility, and a lack of specialized educational features \cite{kuhail2023interacting}. On the other hand, commercial bot-management dashboards introduce further friction for academic use. Similarly, web-based tools such as Dyno that require OAuth2 flows and browser interfaces and are not optimized for mobile use \cite{dynoWebDashboard2025}. At the same time, subscription-oriented services like MEE6 raise concerns about data protection and restrict extensibility through proprietary APIs \cite{mee6Pricing2025}. These constraints are problematic as instructors increasingly rely on mobile devices for on-the-go classroom management and real-time analytics.

This paper presents \textit{InsightEdu}\footnote{GitHub repository: \url{https://github.com/TUM-HN/InsightEdu}}, an open-source iOS application that simplifies Discord-based educational bot management and enables lightweight learning analytics. Building on prior evidence from the \textit{TUMDiscordBot} research prototype—which demonstrated how slash commands can support attendance tracking, tutor-session feedback, and multi-step surveys in introductory programming courses \cite{berrezueta2024interactive}, \textit{InsightEdu} removes the barrier of command-line workflows and JSON-encoded configurations by providing an intuitive and touch-centric interface. Its architecture decouples bot logic from presentation via a RESTful API gateway, \textit{TUMDiscordBot-API}, preserving proven pedagogical functionality while enabling secure, mobile-optimized control.

\textit{InsightEdu} targets three core classroom management tasks with a streamlined mobile experience:
\begin{itemize}
\item \textbf{Attendance tracking}: Initiate, monitor, and record participation.
\item \textbf{Feedback collection}: Collect student feedback about teaching sessions, exercises, tests, and illustrate the results.
\item \textbf{Survey administration}: Create and distribute simple and complex surveys.
\end{itemize}

The remainder of this paper is organized as follows. Section~\ref{RW} surveys educational chatbots and mobile EdTech design. Section~\ref{sec:system_architecture} details the \textit{InsightEdu} system architecture and implementation. Section~\ref{E} describes the study design, tasks, and metrics used to assess technical performance and usability. Section~\ref{R} reports quantitative efficiency, error rates, and latency. Section~\ref{D} reflects on implications for educational technology, architectural lessons, limitations, and future research directions. Finally, Section~\ref{C} concludes the paper.

\section{Related Work}\label{RW}

The integration of educational chatbots has gained traction as a tool to enhance student engagement and collect learning analytics.
Kuhail et al.~\cite{kuhail2023interacting} identified that technical complexity remains the primary barrier to chatbot adoption among educators. While Discord-based bots have proven effective in improving student-instructor interaction in computer science courses, existing solutions often require command-line expertise for deployment and configuration~\cite{berrezueta2024interactive}. Furthermore, traditional Learning Management Systems (LMS) like Moodle Cloud incur substantial annual costs (up to USD 21,360)~\cite{moodleCloudPricing2025}. Discord’s free hosting model provides a cost-effective alternative for budget-constrained institutions.

Modern declarative UI frameworks like SwiftUI have revolutionized educational technology by enabling rapid prototyping with native performance and built-in accessibility~\cite{appleSwiftUIDoc2025, appleSwiftUITut2025}. SwiftData further streamlines this by providing code-first data persistence that integrates with SwiftUI’s reactive data flow, replacing the high boilerplate requirements of legacy frameworks like \textit{Core Data}~\cite{wwdc2023SwiftData, appleCoreData}.

Model-View-ViewModel (MVVM) is the preferred architectural pattern for SwiftUI applications~\cite{appeloper2022stop}. Unlike MVC or VIPER, MVVM’s reactive binding model aligns naturally with declarative syntax, reducing the impedance mismatch between logic and presentation layers~\cite{anderson2012model, aljamea2018mmvmi}.

\section{System Architecture and Implementation}\label{sec:system_architecture}

\textit{InsightEdu} employs a layered architecture that separates presentation logic from bot functionality, allowing for mobile-first management of Discord-based learning activities.

\subsection{Overview and Requirements}
The system comprises three core components: (1) a SwiftUI-based iOS client, (2) a Flask REST API, and (3) a Python Discord bot built with py-cord. 

Beyond documenting instructors’ command-line pain points, such as remembering slash, command syntax, and hand-editing JSON, we also identified a lack of real-time feedback during lectures and a reliance on external tools for post-hoc data processing. We addressed these by refining the functional (FR) and nonfunctional (NFR) requirements in workshops with tutor cohorts and aligning them with capabilities already validated by TUMDiscordBot:

\begin{itemize}
    \item \textbf{FR-1}: Start/stop attendance check.
    \item \textbf{FR-2}: Simple surveys with button/emoji reactions.
    \item \textbf{FR-3}: Complex multi-question surveys with free-text responses.
    \item \textbf{FR-4}: Trigger session feedback dialogs.
    \item \textbf{FR-5}: Persist bot settings (token, development mode) and graphical user interface (GUI) state locally.
    \item \textbf{FR-6}: Display historical survey/attendance data with interactive charts.
    \item \textbf{NFR-1}: GUI actions should complete in $\leq$300\,ms round-trip over campus Wi-Fi;
    \item \textbf{NFR-2}: GUI follows Apple’s Human Interface Guidelines;
    \item \textbf{NFR-3}: All bot-control endpoints require authentication;
    \item \textbf{NFR-4}: Multiple bot instances are supported;
    \item \textbf{NFR-5}: The design anticipates migration to JWT and finer-grained scopes.
\end{itemize}

Our bot setup runs on a Raspberry Pi (4GB RAM) kept on campus. It keeps costs low while handling typical classes of 100–150 students per session.

\subsection{User Interface Design}

The UI is touch-centric and adheres to Apple’s Human Interface Guidelines (HIG)~\cite{appleHIG2024}:

\begin{itemize}
    \item \textit{Hierarchical navigation}: Clear movement between features using \texttt{NavigationStack}.
    \item \textit{Touch-optimized controls}: Large hit targets for mobile use.
    \item \textit{Immediate visual feedback}: Responsive interactions.
    \item \textit{Data visualization}: Charts for feedback/survey summaries (Apple Charts~\cite{appleCharts2024}).
\end{itemize}

\begin{figure}[h]
\centering
\includegraphics[width=0.7\columnwidth]{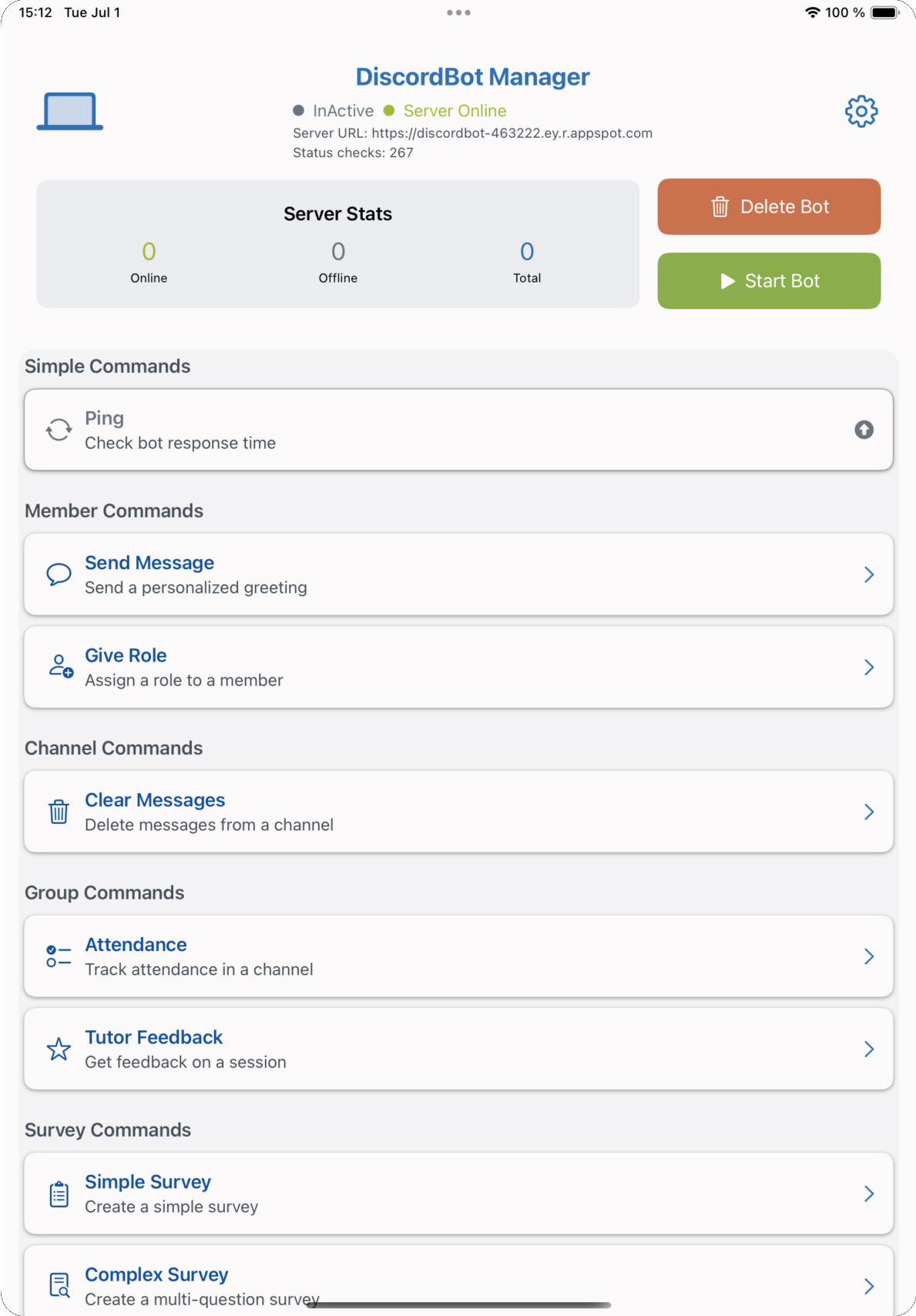}
\caption{Main screen of the \textit{InsightEdu} iOS application.}
\label{fig:main_screen}
\end{figure}

Figure~\ref{fig:main_screen} presents the main screen of the \textit{InsightEdu} iOS application, which provides educators with direct access to educational Discord bot management features. Among its design we have: \\
\\
\textbf{UI Elements - Header Section}
\begin{itemize}
    \item Title, Status Information, and Settings
\end{itemize}
\textbf{UI Elements - Server Controls}
\begin{itemize}
    \item Server Stats, Action Buttons (Delete Bot, Start Bot).
    \end{itemize}
\textbf{UI Elements - Command Categories}
The commands are grouped into functional blocks:
\begin{itemize}
    \item Simple Commands: \textit{Ping:}
    \item Member Commands: \textit{Send Message} and \textit{Give Role} 
    \item Channel Commands: \textit{Clear Messages} 
    \item Group Commands: \textit{Attendance} and \textit{Instructor Feedback} 
    \item Survey Commands: \textit{Simple Survey} and \textit{Complex Survey} 
\end{itemize}

The UI is hierarchical and task-oriented, with large, clearly labeled buttons for quick navigation.
Additionally, a short description accompanies each feature to reduce cognitive load.

\subsection{Server Architecture and REST API}

The server uses Flask~\cite{flask2024} and a dual-threaded design, a REST thread handles HTTP requests and authentication, while a py-cord event loop maintains a persistent Discord Gateway connection.

Flask blueprints organize endpoints by domain, following the \textit{Level 2 REST} conventions. The client communicates via a small, versioned set of JSON endpoints; requests use short-lived bearer tokens or API keys, with RFC~7807-style error payloads. 

Error handling standardizes JSON responses and status codes. Audit logging uses structured JSON with daily rotation; CSV exports (\texttt{data/attendance/*.csv}, \texttt{data/surveys/*.csv}) support institutional analytics and manual inspection.

\subsection{Discord Bot Implementation}

The Discord bot is implemented with py-cord~\cite{pycordDocs2025} and exposes standardized slash commands~\cite{discordSlashCommands2021}. Command categories include:

\paragraph{\textbf{Attendance Tracking}.} Through \textit{InsightEdu}, instructors can start sessions, monitor real-time check-ins, and review history. The workflow proceeds from the session start (via the attendance code for authorization), student check-ins via direct message with this attendance code, and session finalization by the instructor.

\paragraph{\textbf{Feedback Collection.}} \textit{InsightEdu} collects structured feedback (rating scales, open-text) and visualizes aggregates. 

\begin{figure}[h]
\centering
\includegraphics[width=0.65\columnwidth]{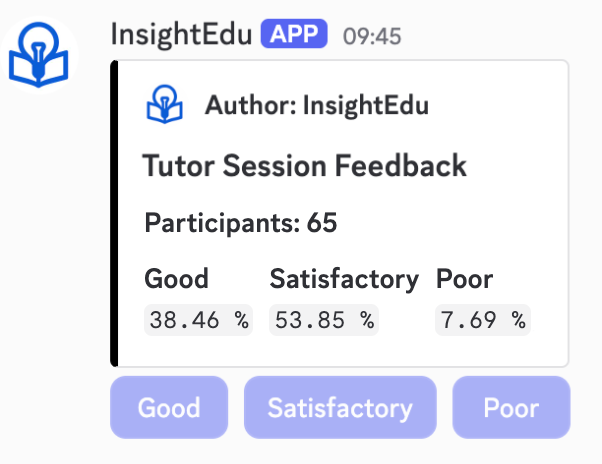}
\caption{Immediate feedback results for students.}
\label{fig:feedback_session}
\end{figure}

The Discord bot also gives students instant feedback. Figure~\ref{fig:feedback_session} shows how their responses are spread across satisfaction levels, making the overall sentiment transparent.

\paragraph{\textbf{Survey Administration.}} Instructors build complex or straightforward surveys, set time limits, and view the results analytics. 

\begin{figure}[h]
\centering
\includegraphics[width=0.7\columnwidth]{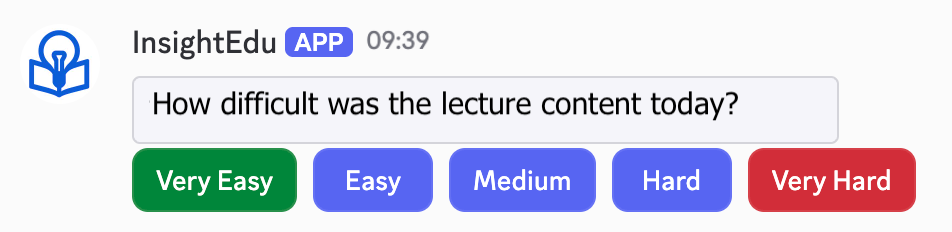}
\caption{Simple survey in a Discord channel.}
\label{fig:simple_survey}
\end{figure}

\begin{figure}[ht]
  \centering
  \includegraphics[width=0.65\linewidth]{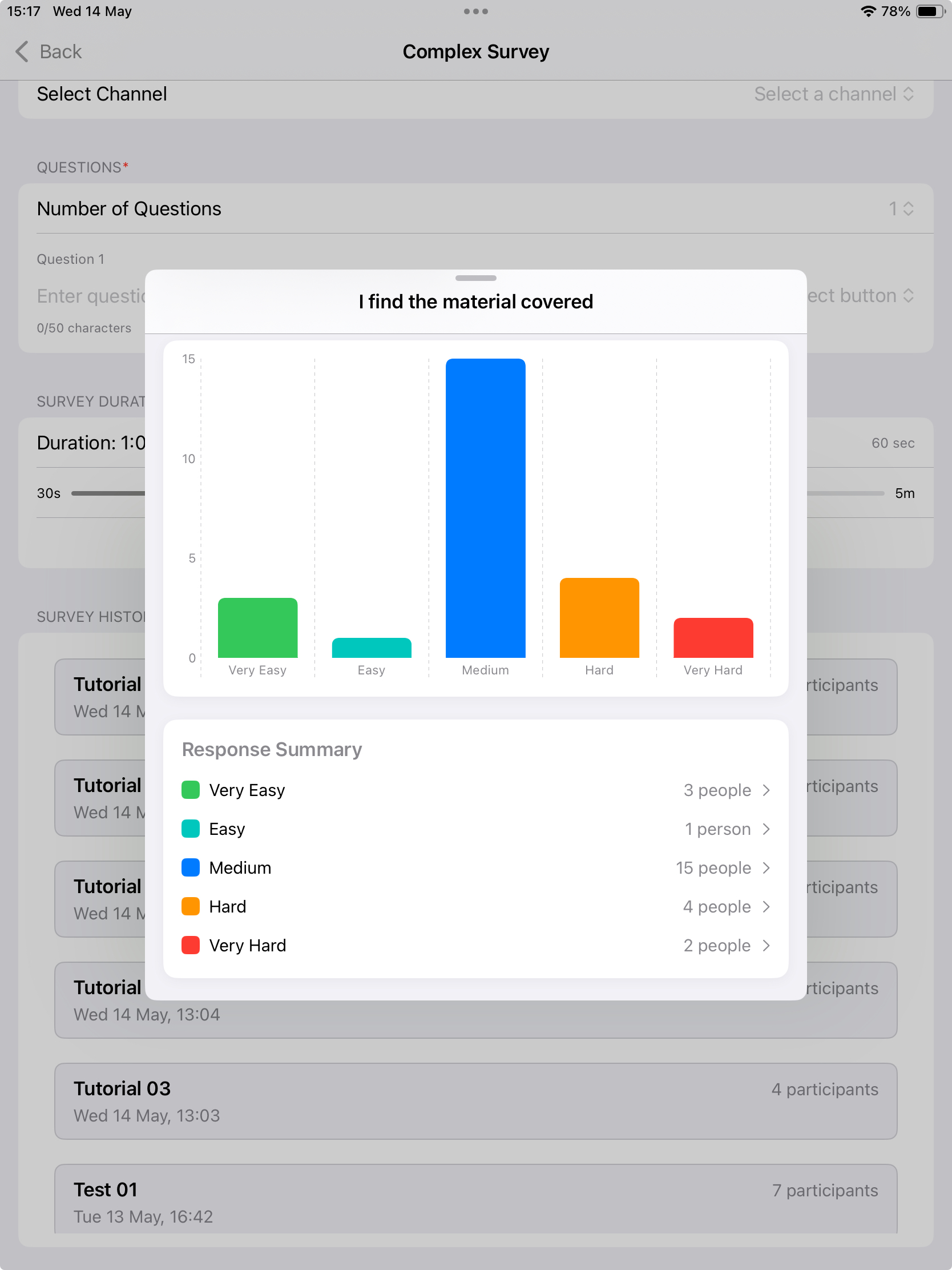}
  \caption{Instructors' results view.}
  \label{fig:materialFeedback}
\end{figure}

The bot offers two types of survey interactions in Discord channels: simple-question surveys and multi-step surveys. Figure~\ref{fig:simple_survey} depicts a five-level difficulty survey suitable for rapid, in-lecture feedback, while Figure~\ref{fig:materialFeedback} presents the instructor’s real-time results.

\begin{figure}[h]
\centering
\includegraphics[width=0.7\columnwidth]{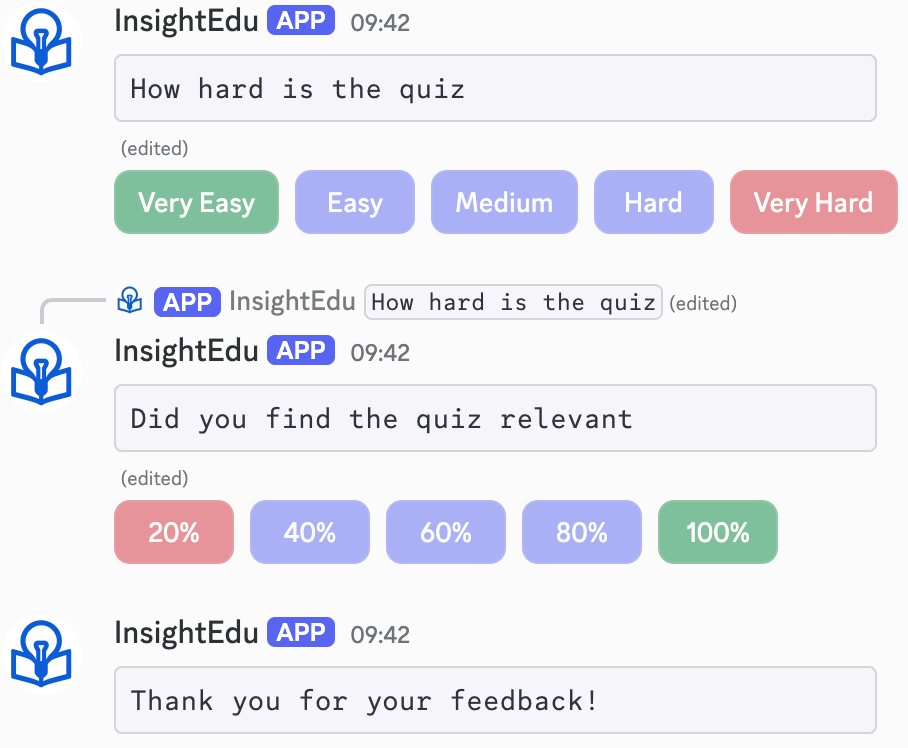}
\caption{Complex survey in a Discord channel.}
\label{fig:complex_survey_dm}
\end{figure}

For more detailed feedback, instructors can create complex surveys that begin in the channel and continue via direct messages after clicking the "\textit{Participate}" button, allowing a sequence of questions (see Figure~\ref{fig:complex_survey_dm}). The results of these complex surveys are also presented to the instructors in similar charts, as shown in Figure~\ref{fig:materialFeedback}. 

\section{Evaluation}\label{E}

The evaluation validated \textit{InsightEdu} through a comparative study contrasting traditional CLI-based bot management with the mobile interface. Twenty experienced instructors with varying technical backgrounds performed six core tasks: attendance tracking, complex and straightforward survey deployment, feedback collection, configuration management, and historical data analysis.

Tests were conducted using a Raspberry Pi 4B over campus Wi-Fi to simulate realistic classroom environments. Usability was assessed using Jakob Nielsen’s heuristics, while cognitive load was measured via the NASA-TLX index. Quantitative metrics included task completion time, error rates, and system response latency. Qualitative assessment captured subjective satisfaction and interface preferences through structured interviews and rankings.

\section{Performance and Usability Results}\label{R}

Quantitative analysis confirms that the \textit{InsightEdu} mobile interface significantly outperforms traditional CLI workflows in efficiency, accuracy, and user satisfaction.

\textbf{Efficiency and Error Rates.}
The mobile interface achieved a 40\% reduction in median task completion time. The most substantial gains occurred in complex survey creation (65\% faster) and data analysis (55\% faster). By eliminating command-line syntax, error rates dropped from an average of 2.3 per task with CLI commands to almost none. While CLI errors were primarily syntax-related, in the app, they were essentially network connectivity issues.

\textbf{Technical Performance.}
The system met all responsiveness targets over the campus Network. Latency metrics are summarized in Table~\ref{tab:latency}. Database operations averaged 180 ms, and Discord API interactions for command execution averaged 250 ms. Analytics visualizations were generated with sub-100 ms latency.

\begin{table}[h]
\caption{System Response Latency Metrics}
\label{tab:latency}
\begin{tabular}{lr}
\toprule
Operation Type & Average Latency \\
\midrule
Local Chart Rendering & $<$ 100 ms \\
Database Operations & 180 ms \\
Discord API Interaction & 250 ms \\
\textbf{Total Round-trip (95th percentile)} & \textbf{$<$ 300 ms} \\
\bottomrule
\end{tabular}
\end{table}

\textbf{Usability and Cognitive Load.}
Heuristic evaluation yielded an average score of 4.2/5.0, with \textit{Aesthetic Design} scoring highest (4.8/5.0). Qualitative feedback showed a 92\% preference for the mobile application. NASA-TLX assessment indicated a significant reduction in perceived mental demand, as the interface moved from recall-based commands to recognition-based touch interactions.

\textbf{Deployment Scalability.}
Stress tests on a Raspberry Pi 4B (4GB RAM) validated the system's low-cost feasibility:

\textit{- Capacity:} Supported 12 concurrent instructors managing 20--50 students each.

\textit{- Resources:} Memory usage stabilized at 1.2GB; CPU utilization averaged 15--20\%.

\textit{- Storage/Network:} Bandwidth remained minimal ($<$1KB per API request), and storage requirements were low ($\approx$50MB per semester).

\textit{- Reliability:} Automatic reconnection restored full functionality within 30--60 seconds after network interruptions.

\section{Discussion}\label{D}

This section analyzes the trade-offs between mobile-first design and the constraints of Discord-mediated instruction.

\subsection{Implications for Educational Technology}

The success of InsightEdu demonstrates that sophisticated learning analytics do not require expensive infrastructure. By combining free Discord hosting with low-cost Raspberry Pi hardware, the barrier to entry for interactive teaching is reduced by orders of magnitude compared to traditional LMS platforms.
The 40\% reduction in task completion time validates the importance of removing command-line friction for adoption among non-technical faculty. InsightEdu allows instructors to pivot from "managing tools" to "responsive pedagogy," using real-time data to adjust lecture pacing.

\subsection{Architectural Lessons and Design Patterns}

While the MVVM pattern and dual-threaded Flask architecture provided high development speed, we identified some tensions:

\textbf{Simplicity vs. Scalability:} The single-process model is easy to deploy but introduces a single point of failure. Future iterations might require containerization for better fault isolation.

\textbf{Security Pragmatism:} Using static API keys (Level 2 REST) was appropriate for this cohort, though institutional scaling will necessitate transitioning to JWT or SSO integration.

\textbf{Platform Synergy:} Leveraging Discord’s reliable WebSocket infrastructure allowed us to focus on UX rather than protocol stability, confirming that "building on proven platforms" is superior to custom-built chat solutions for EdTech.

\subsection{Limitations and Threats to Validity}

The study’s primary limitations include a small participant pool ($n = 20$) and a lack of longitudinal data to track long-term adaptation. Furthermore, the system remains dependent on Discord’s API and Terms of Service. While manageable for pilots, institutional adoption requires contingency planning for platform-wide changes or service outages.

\subsection{Future Research Directions}

The logical evolution of InsightEdu involves three pillars:

\textbf{AI Recommendations:} Using LLMs to transform student feedback into actionable "teaching tips" (e.g., "70\% of students found the last example difficult; suggest a recap").

\textbf{Cross-Platform Support:} Moving toward Flutter or React Native to support Android users.

\textbf{Pedagogical Outcomes:} Shifting focus from "instructor efficiency" to measuring actual student learning gains and engagement levels over a full semester.

\section{Conclusion}\label{C}

\textit{InsightEdu} bridges a critical usability gap by providing a touch-centric mobile interface for managing Discord-based educational bots. By decoupling bot logic from presentation through a native SwiftUI client and Flask API, the system eliminates the friction of command-line workflows. Evaluation results demonstrate a 40\% reduction in task completion time, lower error rates, and a 92\% instructor preference for the mobile interface over traditional CLI methods.
Beyond usability, the successful deployment on low-cost Raspberry Pi hardware confirms that sophisticated learning analytics are viable in resource-constrained environments. By leveraging Discord's free infrastructure, \textit{InsightEdu} offers an accessible alternative to expensive, traditional LMS deployments. This work provides transferable design guidance for the next generation of EdTech: prioritizing mobile-first accessibility and operational simplicity is essential for achieving broad adoption and improving real-time pedagogical responsiveness in diverse academic settings.

\bibliographystyle{unsrtnat}
\bibliography{1_References}

@String{Computing = "Computing" }

@String{Computer = "{IEEE} Computer" }

@String{Springer = "Springer-Verlag" }

@ArtifactSoftware{R,
    title = {R: A Language and Environment for Statistical Computing},
    author = {{R Core Team}},
    organization = {R Foundation for Statistical Computing},
    address = {Vienna, Austria},
    year = {2019},
    url = {https://www.R-project.org/},
}

@inproceedings{appeloper2022stop,
  title={A comparison of architectural patterns for testability and performance quality for iOS mobile applications development},
  author={Magics-Verkman, Hannelore and Zmaranda, Doina Rodica and Gy{\H{o}}r{\"o}di, Cornelia Aurora and Gy{\H{o}}r{\"o}di, Robert-{\c{S}}tefan},
  booktitle={2023 17th International Conference on Engineering of Modern Electric Systems (EMES)},
  pages={1--4},
  year={2023},
  organization={IEEE}
}

@article{kuhail2023interacting,
  author    = {Kuhail, Mohammad A. and Alturki, Nourah and Alramlawi, Shoroq and others},
  title     = {Interacting with educational chatbots: A systematic review},
  journal   = {Education and Information Technologies},
  volume    = {28},
  pages     = {973--1018},
  year      = {2023},
  publisher = {Springer},
  doi       = {10.1007/s10639-022-11177-3},
  url       = {https://doi.org/10.1007/s10639-022-11177-3}
}

@misc{discordSlashCommands2021,
  author       = {{Discord Inc.}},
  title        = {Slash Commands Are Here},
  howpublished = {\url{https://discord.com/blog/slash-commands-are-here}},
  month        = mar,
  year         = {2021},
  note         = {Accessed: 2025-05-12}
}

@misc{pycordDocs2025,
  author       = {{Pycord Development Team}},
  title        = {Pycord Documentation (v2.5.0)},
  howpublished = {\url{https://docs.pycord.dev/en/v2.6.1/}},
  year         = {2025},
  note         = {Accessed: 2025-05-12}
}

@inproceedings{berrezueta2024interactive,
  author    = {Berrezueta-Guzman, Santiago and Parmacli, Ivan and Krusche, Stephan and Wagner, Stefan},
  title     = {Interactive Learning in Computer Science Education Supported by a Discord Chatbot},
  booktitle = {2024 IEEE 3rd German Education Conference (GECon)},
  year      = {2024},
  pages     = {1--6},
  publisher = {IEEE},
  doi       = {10.1109/GECon62014.2024.10734012},
  url       = {https://doi.org/10.1109/GECon62014.2024.10734012}
}

@article{dynoWebDashboard2025,
  title={Discord Unveiled: A Comprehensive Dataset of Public Communication (2015-2024)},
  author={Aquino, Yan and Bento, Pedro and Buzelin, Arthur and Dayrell, Lucas and Malaquias, Samira and Santana, Caio and Estanislau, Victoria and Dutenhefner, Pedro and Evangelista, Guilherme HG and Porf{\'\i}rio, Luisa G and others},
  journal={arXiv preprint arXiv:2502.00627},
  year={2025}
}

@misc{mee6Pricing2025,
  title        = {MEE6 Wiki},
  howpublished = {\url{https://wiki.mee6.xyz/}},
  note         = {Accessed: 2025-11-07},
  author       = {{MEE6 Community}}
}

@misc{appleSwiftUIDoc2025,
  author       = {{Apple Inc.}},
  title        = {SwiftUI Documentation},
  howpublished = {\url{https://developer.apple.com/documentation/swiftui/}},
  year         = {2025},
  note         = {Accessed: 2025-05-12}
}

@misc{appleSwiftUITut2025,
  author       = {{Apple Inc.}},
  title        = {SwiftUI Tutorials},
  howpublished = {\url{https://developer.apple.com/tutorials/swiftui/}},
  year         = {2025},
  note         = {Accessed: 2025-05-12}
}

@misc{wwdc2023SwiftData,
  author       = {{Apple Inc.}},
  title        = {Meet SwiftData (WWDC 2023)},
  howpublished = {\url{https://developer.apple.com/videos/play/wwdc2023/10187/}},
  year         = {2023},
  note         = {Accessed: 2025-05-12}
}

@misc{appleHIG2024,
  author       = {{Apple Inc.}},
  title        = {Human Interface Guidelines},
  howpublished = {\url{https://developer.apple.com/design/human-interface-guidelines/}},
  year         = {2024},
  note         = {Accessed: 2025-05-14}
}

@misc{appleCharts2024,
  author       = {{Apple Inc.}},
  title        = {Swift Charts Documentation},
  howpublished = {\url{https://developer.apple.com/documentation/charts}},
  year         = {2024},
  note         = {Accessed: 2025-05-14}
}

@misc{flask2024,
  author       = {{Pallets Project}},
  title        = {Flask Documentation},
  howpublished = {\url{https://flask.palletsprojects.com/en/2.3.x/}},
  year         = {2024},
  note         = {Accessed: 2025-05-14}
}

@misc{moodleCloudPricing2025,
  author       = "{MoodleCloud}",
  title        = "{MoodleCloud Standard Plans}",
  howpublished = "\url{https://www.moodlecloud.com/standard-plans/}",
  year         = "2025",
  note         = "[Online; accessed 14-May-2025]"
}

@article{moodle,
  title={A systematic review on trends in using Moodle for teaching and learning},
  author={Gamage, Sithara HPW and Ayres, Jennifer R and Behrend, Monica B},
  journal={International journal of STEM education},
  volume={9},
  number={1},
  pages={9},
  year={2022},
  publisher={Springer}
}

@inproceedings{interactiveLearning,
  author    = {Krusche, Stephan and Berrezueta-Guzman, Jonnathan},
  title     = {Introduction to Programming using Interactive Learning},
  booktitle = {2023 IEEE 35th International Conference on Software Engineering Education and Training (CSEE\&T)},
  year      = {2023},
  pages     = {178--182},
  publisher = {IEEE},
  doi       = {10.1109/CSEET58097.2023.00037},
  url       = {https://doi.org/10.1109/CSEET58097.2023.00037}
}

@inproceedings{wulanjani2018discord,
  author    = {Wulanjani, Arum Nisma},
  title     = {Discord application: Turning a voice chat application for gamers into a virtual listening class},
  booktitle = {Proceedings of the English Language and Literature International Conference (ELLiC)},
  volume    = {2},
  pages     = {115--119},
  year      = {2018},
  publisher = {Universitas Muhammadiyah Semarang},
  address   = {Semarang, Indonesia},
  url       = {https://jurnal.unimus.ac.id/index.php/ELLIC/article/view/3500},
  note      = {Accessed: 2025-09-02}
}

@misc{appleCoreData,
  author       = {{Apple Inc.}},
  title        = {Core Data | Apple Developer Documentation},
  howpublished = {\url{https://developer.apple.com/documentation/coredata}},
  note         = {Accessed: 2025-09-01},
  year         = {2025}
}

@incollection{anderson2012model,
  title={The model-view-viewmodel (mvvm) design pattern},
  author={Anderson, Chris},
  booktitle={Pro Business Applications with Silverlight 5},
  pages={461--499},
  year={2012},
  publisher={Springer}
}

@article{aljamea2018mmvmi,
  title={MMVMi: A Validation Model for MVC and MVVM Design Patterns in iOS Applications.},
  author={Aljamea, Mariam and Alkandari, Mohammad},
  journal={IAENG International Journal of Computer Science},
  volume={45},
  number={3},
  year={2018}
}

@misc{andree2025softskills,
      title={How Soft Skills Shape First-Year Success in Higher Education}, 
      author={Kerstin Andree and Santiago Berrezueta-Guzman and Stephan Krusche and Luise Pufahl and Stefan Wagner},
      year={2025},
      eprint={2505.21696},
      archivePrefix={arXiv},
      primaryClass={cs.CY},
      url={https://arxiv.org/abs/2505.21696}, 
}

\end{document}